\documentclass[twocolumn,prb,showpacs,preprintnumbers,amsmath,amssymb,floatfix,superscriptaddress]{revtex4}
\usepackage{graphicx}
\usepackage{amsmath}
\usepackage{amssymb}
\usepackage{color}

\begin{document}

\title{Percolation-to-hopping crossover in conductor-insulator composites}

\author{G. Ambrosetti}\email{gianluca.ambrosetti@a3.epfl.ch}\affiliation{LPM, Ecole Polytechnique F\'ed\'erale de Lausanne, Station 17, CH-1015 Lausanne, Switzerland}
\author{I. Balberg}\affiliation{The Racah Institute of Physics, The Hebrew University, Jerusalem 91904, Israel}
\author{C. Grimaldi}\email{claudio.grimaldi@epfl.ch}\affiliation{LPM, Ecole Polytechnique F\'ed\'erale de
Lausanne, Station 17, CH-1015 Lausanne, Switzerland}



\begin{abstract}
Here, we show that the conductivity of conductor-insulator composites in which electrons can tunnel from each
conducting particle to all others may display both percolation and tunneling ({\it i.e.}, hopping) regimes depending
on few characteristics of the composite. Specifically, we find that the relevant parameters that give rise to
one regime or the other are $D/\xi$ (where $D$ is the size of the conducting particles and $\xi$ is the
tunneling length) and the specific composite microstructure.
For large values of $D/\xi$, percolation arises when the composite microstructure can be modeled as a
regular lattice that is fractionally occupied by conducting particle, while the tunneling regime
is always obtained for equilibrium distributions of conducting particles in a continuum insulating matrix.
As $D/\xi$ decreases the percolating behavior of the conductivity of lattice-like composites
gradually crosses over to the tunneling-like regime characterizing particle dispersions in the continuum.
For $D/\xi$ values lower than $D/\xi\simeq 5$ the conductivity has tunneling-like behavior independent of
the specific microstructure of the composite.
\end{abstract}
\maketitle

\section{Introduction}
\label{intro}

The conductivity of a conductor-insulator composite material is characterized by a
strong dependence on the volume fraction $\phi$ of the conducting phase present in the system, and
is generally understood as a percolation phenomenon arising from the electrical
connectivity between neighboring or adjacent conducting particles. Specifically,
percolation theory considers the conducting particles as either electrically connected,
with some finite inter-particle conductance, or disconnected.\cite{Stauffer1994,Sahimi2003}
The introduction of this sharp cut-off implies then that below a specific fraction $\phi_c$
(the percolation threshold) the conductivity is zero because there is no sample-spanning
network of connected particles, while for $\phi>\phi_c$ such network is formed and the conductivity
increases as $\simeq (\phi-\phi_c)^t$, where $t$ is a critical exponent.

It is easy to see that the notion of a sharp cut-off applies well to composites
made of large (of the order of one micron or more) conducting particles,
for which two particles can be considered electrically connected only if they essentially touch each other,
it is less clear for the case in which the conducting particles have sizes limited to
a few nanometers.\cite{Balberg2009} In that situation even if the particles do not physically
touch each other, their mean separation in the composite is such that
electrons can still flow via tunneling processes from one particle to the other.
The resulting tunneling conductance decays exponentially with the inter-particle distance
over a characteristic tunneling length $\xi$ which is of the order of
a fraction to a few nanometers depending on the material properties.
Since the tunneling decay does not imply any sharp cut-off, the basic assumption of percolation theory
in describing composites made of nanometric conducting fillers is not justified a priori.
In fact, it turns out that nanocomposites in the dielectric regime, for which the conducting particles
do not touch each other, are better explained by inter-particle tunneling, with no
imposed sharp connectivity criterion, than by the classical percolation
theory.\cite{Ambrosetti2009,Ambrosetti2010} The resulting conductivities of these
systems follow therefore hopping-like (or tunneling-like)
behaviors,\cite{Shklovskii1984,Ambegaokar1971,Seager1974,Overhof1989} with no critical
percolation thresholds as a function of $\phi$.

The possibility of having percolation-like or tunneling-like regimes depending on the size $D$ of
the conducting particles compared to the tunneling length $\xi$, as suggested above and
in Ref.[\onlinecite{Balberg2009}], does not appear to have been further elaborated
in the literature, despite of its relevance to the understanding of transport properties in
conductor-insulator composites.
In the present paper we address this issue by considering a global tunneling network (GTN)
model of conductor-insulator composites,\cite{Ambrosetti2010} where each conducting particle is connected to all
others via tunneling processes. We show that this model permits to treat the
percolation and the tunneling regimes on equal footing, and that one can switch from one
regime to the other depending not only on $D/\xi$ but, most notably, also on the specific distribution
of the conducting phase in the composite.
We illustrate this behavior by considering two idealized realizations of
conductor-insulator composites: a lattice model, where conducting spherical particles of diameter $D$ occupy
randomly a fraction of the sites of a regular lattice, and a continuum model, where the conducting
particles are dispersed with an equilibrium distribution in a continuous insulating medium.
By using both the effective medium approximation (EMA) and Monte Carlo (MC) calculations, we find that
the conductivity of the continuum model has a tunneling-like behavior independent of
the value of $D/\xi$, while the lattice model displays a percolation behavior of the conductivity
only for sufficiently large $D/\xi$ values, which gradually crosses over to a tunneling-like regime as
$D/\xi$ decreases. For $D/\xi\lesssim 5$ we show that
the conductivity of the lattice model is basically indistinguishable from that of the continuum case,
establishing therefore a crossover point from percolation to tunneling behaviors.

This paper is organized as follows. In Sec.~\ref{EMA} we introduce the GTN model, we formulate
the EMA for both lattice and continuum models and we calculate the resulting EMA conductances.
In Sec.~\ref{MC} we present our MC calculations for both the lattice and continuum models, and in
Sec.~\ref{crossover} we discuss the crossover between percolation and tunneling regimes. Discussions and
conclusions are given in Sec.~\ref{concl}

\section{Model and effective medium approximation}
\label{EMA}

The GTN model for conductor-insulator composites is defined by considering
$n$ identical conducting particles contained within a volume $V$. This defines
the volume fraction $\phi=\rho v$ of the conducting phase, where $\rho=n/V$ is the particle density
and $v$ is the volume of a single particle. For the purpose of the present work we limit
the analysis to the relatively simple case of spherical particles of identical diameter $D$, so that $v=\pi D^3/6$.
Next, we assume that the tunneling conductance between any two particles centered at
${\bf r}_i$ and ${\bf r}_j$ is given by:
\begin{equation}
\label{tunnel1}
g(r_{ij})=g_0\exp\!\left[-\frac{2(r_{ij}-D)}{\xi}\right],
\end{equation}
where $g_0$ is a constant ``contact" conductance which in the following we shall
set equal to unity, $\xi$ is the characteristic tunneling length, and
$r_{ij}=\mid\!{\bf r}_i-{\bf r}_j\!\mid$ is the distance between two sphere centers.
For simplicity we further assume that selective tunneling mechanisms arising from excitation energies
can be safely neglected (as in the case for nearest neighbor hopping\cite{Shklovskii1984}).

The set of all tunneling conductances of Eq.~\eqref{tunnel1} defines a resistor network whose
conductivity depends on $\xi$ and $D$, as well as on the volume fraction $\phi$ and on the specific
distribution of the particle centers. As shown below, it turns out that all these
dependencies are well captured by the (single bond) EMA applied to the tunneling resistor
network.\cite{Sahimi2003} This is given by the solution of the following equation:
\begin{equation}
\label{EMA1}
\left\langle\sum_{i\neq j}\frac{g(r_{ij})-\bar{g}}{g(r_{ij})+[Z_iZ_j/(Z_i+Z_j)-1]\bar{g}}\right\rangle=0,
\end{equation}
where $\bar{g}$ is the effective bond conductance and $i$ and $j$ run over the positions of the $n$ particles.
In the above expression, the symbol $\langle\ldots\rangle$ denotes the statistical average over all realizations
of the $n$-particle system. Furthermore, $Z_i$ and $Z_j$ are the coordination numbers of the two ends
of a resistor $g(r_{ij})$, \textit{i.e.}, given a particle at $i$ ($j$), $Z_i$ ($Z_j$) is the number of
particles that are electrically connected to $i$ ($j$). Since it is assumed that all particles are connected to each
other regardless of their relative distances, $Z_i=Z_j=n-1$. In this way, Eq.~\eqref{EMA1} reduces to
\begin{equation}
\label{EMA2}
\left\langle\sum_{i\neq j}\frac{g(r_{ij})-\bar{g}}{g(r_{ij})+[(n-1)/2-1]\bar{g}}\right\rangle=0.
\end{equation}
By noticing that $\sum_{i\neq j}=n(n-1)$ and by using Eq.~\eqref{tunnel1}, we can write Eq.~\eqref{EMA2} more
conveniently as
\begin{equation}
\label{EMA3}
\left\langle\frac{1}{n}\sum_{i\neq j}\frac{1}{g^*\exp[2(r_{ij}-D)/\xi]+1}\right\rangle=2,
\end{equation}
where for large $n$ the dimensionless conductance $g^*=[(n-1)/2-1]\bar{g}/g_0\simeq n\bar{g}/2g_0$ coincides
with the conductance between two random nodes of the network with all bond conductances equal to $\bar{g}/g_0$.\cite{Lopez2006}

By multiplying each term of the summation over $i,j$ by $\int\! d{\bf r}\, \delta({\bf r}-{\bf r}_{ij})=1$,
where the integration is extended over the entire volume $V$,
the left-hand side of Eq.~\eqref{EMA3} becomes
\begin{align}
\label{EMA4}
&\left\langle\frac{1}{n}\sum_{i\neq j}\int\! d{\bf r}\,\delta({\bf r}-{\bf r}_{ij})\frac{1}{g^*\exp[2(r_{ij}-D)/\xi]+1}\right\rangle\nonumber \\
&=\int\! d{\bf r}\left\langle\frac{1}{n}\sum_{i\neq j}\delta({\bf r}-{\bf r}_{ij})\right\rangle
\frac{1}{g^*\exp[2(r-D)/\xi]+1}.
\end{align}
The above expression can be rewritten in terms of the pair distribution function $g_2({\bf r})$ by
noticing that, by definition:\cite{Hansen}
\begin{equation}
\label{pdf}
\rho g_2({\bf r})=\left\langle\frac{1}{n}\sum_{i\neq j}\delta({\bf r}-{\bf r}_{ij})\right\rangle ,
\end{equation}
so that Eq.~\eqref{EMA3} reduces to
\begin{equation}
\label{EMA5}
\int_0^\infty\! dr \frac{4\pi r^2\rho\, g_2(r)}{g^*\exp[2(r-D)/\xi]+1}=2,
\end{equation}
where $g_2(r)=\int\! d\Omega\, g_2({\bf r})/4\pi$ is the radial distribution function (rdf)
and the upper limit of the integration has been set to infinity because of the exponential
decay of the integrand. Note that Eq.~\ref{EMA5} is similar to the result obtained
in Ref.~\onlinecite{Movaghar1981}.

An alternative and useful version of Eq.~\eqref{EMA5} can be obtained by introducing the function
\begin{equation}
\label{W}
W(r)= \frac{1}{g^*\exp[2(r-D)/\xi]+1}=\frac{1}{\exp[2(r-r^*)/\xi]+1},
\end{equation}
with
\begin{equation}
\label{rstar}
r^*=D+\frac{\xi}{2}\ln\!\left(\frac{1}{g^*}\right)
\end{equation}
and integrating Eq.~\eqref{EMA5} by parts:
\begin{equation}
\label{EMA6}
\int_0^\infty\! dr Z(r) \left(-\frac{dW(r)}{dr}\right)=2,
\end{equation}
where
\begin{equation}
\label{zeta}
Z(r)=\int_0^r\! dr' 4\pi r'^2 \rho\,g_2(r'),
\end{equation}
is the cumulative coordination number function ({\it i.e.}, the number of
spheres whose centers are within a distance $r$ from the center of a given sphere).

\begin{figure}[t!]
\begin{center}
\includegraphics[scale=0.3,clip=true]{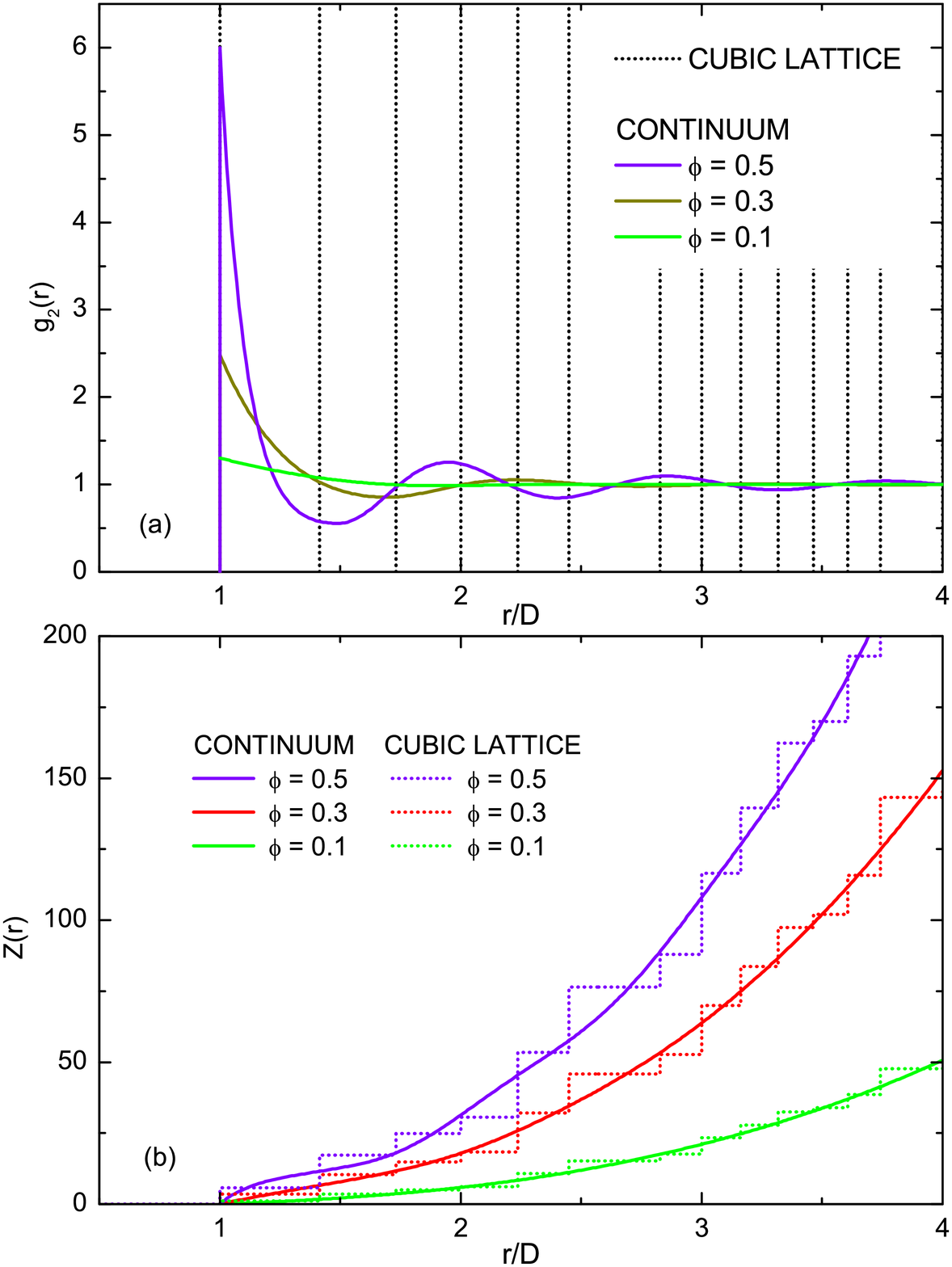}
\caption{(Color online) (a) Radial distribution function $g_2(r)$ for the lattice (dotted lines) and
the continuum (solid lines) models. For the lattice model the vertical dotted lines indicate the positions
of the delta-peaks centered at $R_k$ for $k=1,2,3,\ldots$. The $g_2(r)$ curves for different $\phi$ values
of the continuum model have been obtained by using the results of Ref.~\onlinecite{Trokhymchuk2005}.
(b) Cumulative coordination number $Z(r)$, Eq.~\eqref{zeta}, for the lattice (dotted lines) and the continuum (solid lines)
models for different $\phi$ values.  }\label{fig1}
\end{center}
\end{figure}

\subsection{Lattice and continuum models}
\label{models}

As it is clear from the EMA equations \eqref{EMA5} and \eqref{EMA6}, all the informations on the
spatial distribution of the conducting particles is contained in the rdf
$g_2(r)$, whose dependencies on $r$ and $\rho$ govern the behavior of the EMA
conductance $g^*$. Here, we consider two possible realizations of a conductor-insulator composite
which corresponds to two rather extreme forms of $g_2(r)$.

In the first case, we consider a simple cubic lattice with a lattice constant equal to the
sphere diameter $D$, where only a fraction $p$ of the lattice sites is occupied, randomly, by
the spherical conducting particles, while the remaining fraction $1-p$ is occupied by insulating spheres
of equal diameter (as in the Scher and Zallen model of Ref.~\onlinecite{Scher1970}). The $i$-th conducting particle occupies therefore the position ${\bf r}_i={\bf R}$
with probability $p=n/N$, where ${\bf R}$ is the direct lattice vector running over all $N$ sites of the cube.
In this way, the pair distribution function defined in Eq.~\eqref{pdf} reduces to
\begin{equation}
\label{rdf2}
\rho g_2({\bf r})=p\sum_{{\bf R}\neq {\bf 0}}\delta({\bf r}-{\bf R}),
\end{equation}
which leads to an rdf of the form:
\begin{equation}
\label{rdf3}
\rho g_2(r)=\frac{p}{4\pi}\sum_{k=1,2,\ldots}\frac{N_k}{R_k^2}\,\delta(r-R_k),
\end{equation}
where $N_k$ is the number of the $k$-th nearest neighbors being at distance $R_k$ from
a reference particle set at the origin. The characteristic feature of this fractionally
occupied lattice model is therefore that its rdf,
shown in Fig.~\ref{fig1}(a) by dotted lines, is given
by delta-peaks whose positions do not change with $p$. As we shall see in the following, this
feature is directly related to the appearance of a percolation behavior of the EMA conductance $g^*$
for large $D/\xi$ values.

The second model of particle distribution is given by an equilibrated dispersion of impenetrable
spheres in the continuum. Typical examples of the resulting $g_2(r)$ for this case, obtained by
using the formula provided in Ref.~\onlinecite{Trokhymchuk2005}, are shown in
Fig.~\ref{fig1}(a) by solid lines and for different values of the volume fraction $\phi$.
As opposed to the lattice case, the rdf of an equilibrium dispersion of hard spheres is continuous and
nonzero in the whole $r\ge D$ range, which we anticipate here to be the pre-requisite for an
hopping behavior of the composite conductivity.

In Fig.~\ref{fig1}(b) we compare the coordination number function $Z(r)$, Eq.~\eqref{zeta},
for the lattice (dotted lines) and continuum (solid lines) models. Contrary to the continuum model
in which $Z(r)$ increases smoothly with $r$, the lattice $Z(r)$ has a step-like increase. At large $r$
however the main contribution of the lattice $Z(r)$ goes as $(4/3)\pi\rho r^3$, meaning that
the distribution of spheres in a sparsely occupied lattice is basically that of point particles in the continuum.

\subsection{EMA conductance}
\label{EMAcond}

\begin{figure*}[t]
\begin{center}
\includegraphics[scale=0.64,clip=true]{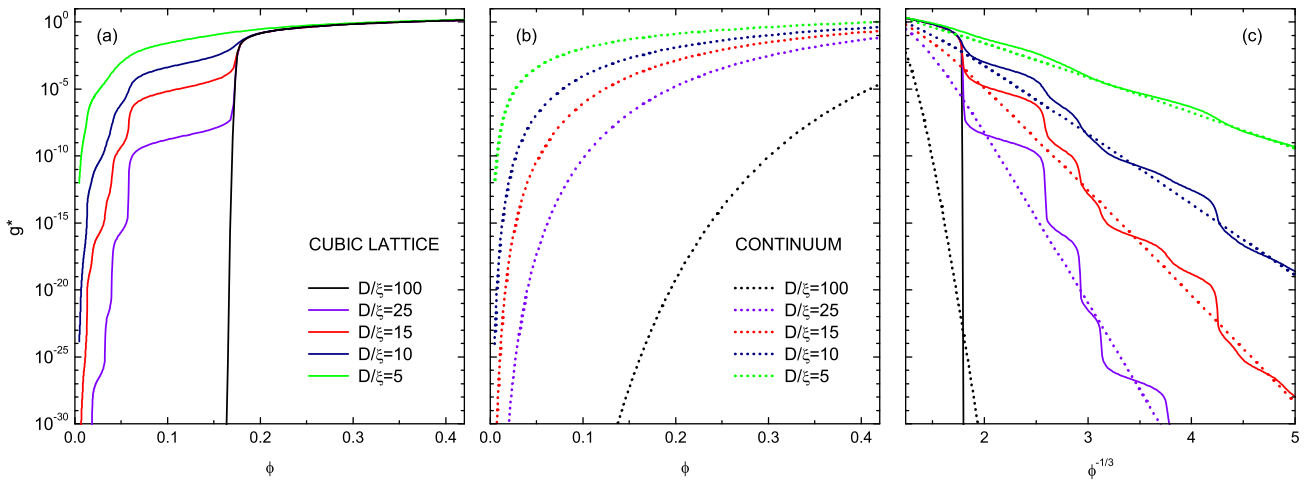}
\caption{(Color online) EMA conductance $g^*$ for (a) the cubic lattice model and (b) the continuum model for different values
of $D/\xi$. (c) EMA conductance of (a) and (b) plotted as a function of $\phi^{-1/3}$.  }\label{fig2}
\end{center}
\end{figure*}

Let us start by considering the solution of the EMA equation for the
fractionally occupied cubic lattice case. By using Eq.~\eqref{rdf3} we find that
Eq.~\eqref{EMA5} reduces to:
\begin{equation}
\label{EMAcube1}
p\sum_{k=1,2,\ldots}\frac{N_k}{g^*\exp[2(R_k-D)/\xi]+1}=2.
\end{equation}
In the limiting case of very large particle sizes such that $D/\xi\rightarrow\infty$, only the first ($k=1$) nearest
neighbors with $R_1=D$ and $N_1=6$ contribute to the summation in Eq.~\eqref{EMAcube1}, which reduces
to $6p/(g^*+1)=2$. Hence, in this limit, the resulting conductivity $g^*=3(p-1/3)$ has a percolation behavior
and vanishes at the critical fraction $p_1=1/3$.
For finite $D/\xi$ values, the next nearest neighbors also contribute to
the network conductivity. For example, by assuming that $D/\xi$ is large enough to retain
only the terms up to the second nearest neighbors
($N_2=12$ and $R_2=\sqrt{2}D$) we get that Eq.~\eqref{EMAcube1} becomes
\begin{equation}
\label{EMAcube2}
\frac{6p}{g^*+1}+\frac{12p}{g^*\exp[2D(\sqrt{2}-1)/\xi]+1}=2,
\end{equation}
whose solution still behaves as $g^*\simeq p-p_1$ for $p>p_1$ but remains finite,
albeit exponentially small,
at lower $p$ values. This is  because the tunneling to the $2$-nd nearest neighbors
vanishes only at $p=p_2=1/9$. Considering the whole set of neighbors when $p$ decreases, one finds
that the solution of
Eq.~\eqref{EMAcube1} becomes characterized by a monotonous decrease of $g^*$ punctuated by sharp
(for small $D/\xi$ values) drops at  $p=p_k$ with  $p_k=2/(\sum_{k'=1}^k N_{k'})$. Furthermore, in the vicinity
of each $p_k$, the conductance follows the EMA power law behavior $\simeq (p-p_k)$ for $p>p_k$.

This feature is illustrated in Fig.~\ref{fig2}(a) where we show numerical solutions of Eq.~\eqref{EMAcube1}
for different values of $D/\xi$.
For $D/\xi=100$, the first percolation transition at $p_1=1/3$
($\phi_1=p_1\pi/6\simeq 0.175$) causes a drop of the conductivity of about $36$ orders of magnitude
(not shown in the figure) compared to the conductivity at the close packing fraction $\pi/6$
that corresponds to the lattice sites that are all occupied by the conducting spheres.
Such drop of the conductivity is well beyond the maximum
range of conductivity values measured in real composites,\cite{notespan} and so, for all practical purposes, the
$D/\xi=100$ case behaves as $(\phi-\phi_1)^t$ with $t=1$ being the EMA critical exponent.
For $D/\xi=25$ the first transition at $\phi_1$ leads to a drop of $g^*$ of only $8$ orders of magnitude,
and the transitions at lower values of $\phi$ are clearly visible in the figure. For lower values of $D/\xi$,
the drops of $g^*$ are further reduced and the transitions are much smoother, due of course to the fact
that for these $D/\xi$ values the probability of tunneling to neighbors that are farther apart is enhanced.
For $D/\xi=5$ the variation of $g^*$ is so smooth in the whole range of $\phi$ that the underlying
lattice structure can be considered as completely irrelevant.

In contrast with the lattice case, the EMA conductance $g^*$ resulting from an equilibrium distribution
of conducting spheres in the continuum does not display any percolation behavior even for large $D/\xi$
values. This is shown in Fig.~\ref{fig2}(b) where solutions of Eq.~\eqref{EMA5}, with $g_2(r)$ as
given in Ref.~\onlinecite{Trokhymchuk2005}, are plotted for the same $D/\xi$ values of Fig.~\ref{fig1}(a).
The lack of percolation behavior in this case is due to the fact that the corresponding rdf [solid lines in
Fig.~\ref{fig1}(a)] is always nonzero for $r\ge D$ and it does not vary much even for $\phi$ values
close to the packing fraction of the simple cubic lattice ($\phi=\pi/6\simeq 0.524$).
To see how the corresponding tunneling behavior arises, let us consider the EMA in the form of Eq.~\eqref{EMA6}.
Since, as shown in Fig.~\ref{fig1}(b),
the coordination number function $Z(r)$ is a smooth increasing function of $r/D$ and, given that
$-dW/dr$ is peaked at $r=r^*$ with spread $\simeq\xi$,\cite{noteW} we can approximate Eq.~\eqref{EMA6} by
\begin{equation}
\label{EMA7}
Z(r^*)=2,
\end{equation}
so that, by applying Eq.~\eqref{rstar}, the EMA conductance becomes
\begin{equation}
\label{EMA8}
g^*=\exp\left[-\frac{2(r^*-D)}{\xi}\right]
\end{equation}
where $r^*$ is such that Eq.~\eqref{EMA7} is satisfied. In passing, we note that
Eqs.~\eqref{EMA7} and \eqref{EMA8} represent the EMA equivalent of the critical path
approximation (CPA) of Refs.~\onlinecite{Seager1974,Overhof1989}.\cite{noteCPA}
For sufficiently large $r^*$ ({\it i.e.}, small $g^*$) the coordination number goes
as $Z(r^*)\simeq (4/3)\pi \rho(r^*)^3$ and from Eq.~\eqref{EMA8} we obtain that
\begin{equation}
\label{EMA9}
g^*\simeq \exp\left[-\frac{2D}{\xi}\left(\frac{0.63}{\phi^{1/3}}-1\right)\right],
\end{equation}
which coincides, if the coefficient $0.63$ in the exponent is replaced by $0.7$, with the low-density
hopping behavior as obtained from the CPA.\cite{Seager1974} In the same low-density limit, Eq.~\eqref{EMA9}
(with a slightly different coefficient in the exponent) has been derived also in Ref.~\onlinecite{Movaghar1981} by
using an EMA-based procedure similar to the one presented here.

\begin{figure*}[t]
\begin{center}
\includegraphics[scale=0.64,clip=true]{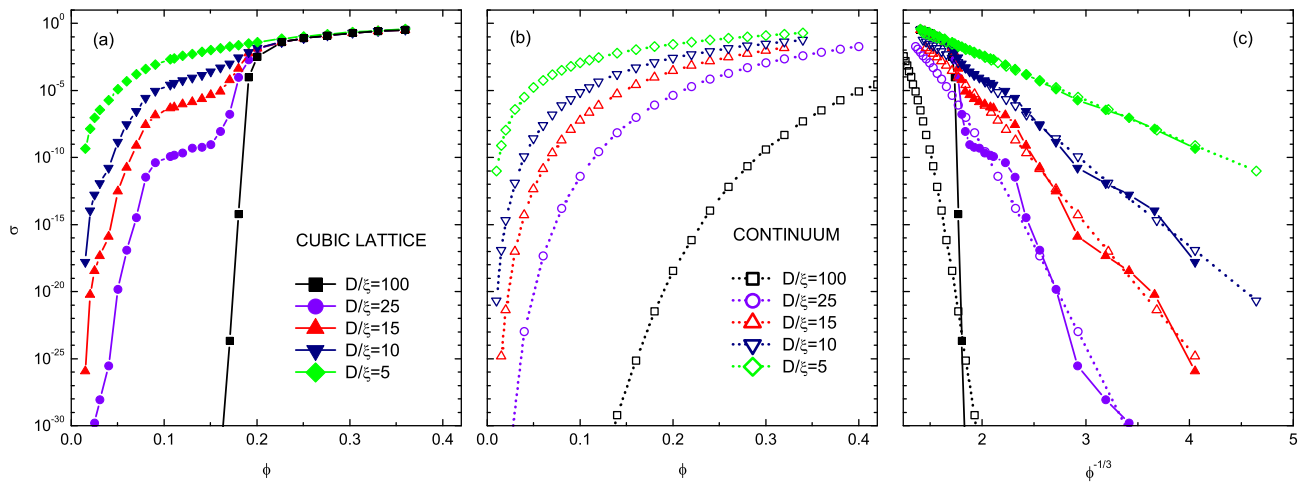}
\caption{(Color online) Monte Carlo conductance  for (a) the cubic lattice model and (b) the continuum model for different values
of $D/\xi$. (c) The same results of (a) and (b) plotted as a function of $\phi^{-1/3}$. }\label{fig3}
\end{center}
\end{figure*}

The hopping-like (or tunneling-like) dependence of $g^*$ for the continuum model is further
illustrated in Fig.~\ref{fig2}(c) where the results of Fig.~\ref{fig2}(b)
have been re-plotted as a function of $\phi^{-1/3}$ (dotted lines).  In Fig.~\ref{fig2}(c) we show for
comparison also the $g^*$ versus $\phi^{-1/3}$ curves (solid lines) of the lattice model results
of Fig.~\ref{fig2}(a), which tend, with the decrease of $D/\xi$, to the tunneling behavior of the continuum
model. An asymptotic equivalence between the two models is indeed expected because, as noted above, the underlying
microstructures become irrelevant for small $D/\xi$. Mathematically, this can be seen from Eq.~\eqref{EMA6}
since $-dW/dr$ averages $Z(r)$ within a distance $\simeq\xi$ around $r^*$, and thus the differences between the lattice
and the continuum models are blurred for sufficiently large $\xi$.  Note also that for the lattice model
$Z(r^*)\simeq (4/3)\pi \rho (r^*)^3$ for large $r^*$, implying that  $g^*$ tends asymptotically to
the results given by Eq.~\eqref{EMA9}, as shown by the solid lines in Fig.~\ref{fig2}(c).

\section{Monte Carlo results}
\label{MC}

Let us turn now to a comparison of our EMA results of the previous section with full MC
calculations of the conductivity for both the lattice and continuum models.
For the continuum we have generated equilibrium distributions of impenetrable spheres
inside a cubic cell with periodic boundary conditions following the procedure that we outlined
previously in Refs.~\onlinecite{Ambrosetti2009,Ambrosetti2010}. For the lattice model, we simply
populated a given fraction of the cubic
lattice sites with conducting spheres with diameter equal to the cubic lattice constant.

To calculate the conductivity resulting from the lattice and continuum models, we ascribed to each pair
of particles the tunneling conductance of Eq.~\eqref{tunnel1}, and performed a numerical decimation
of the resulting resistor network.\cite{Ambrosetti2009,Ambrosetti2010,Johner2008} To reduce computational times of the decimation procedure we introduced an artificial maximum distance between the particles beyond which the resulting bond conductance can be safely neglected.\cite{Ambrosetti2009,Ambrosetti2010}

The results of the calculated conductivity are plotted in Fig.~\ref{fig3} for the same parameter
values that were used for Fig.~\ref{fig2}.
Each symbol is the outcome of $N_R=50$ realizations of systems of $N_P\sim1000$ spheres.
The logarithm average of the results was considered since, due to the exponential dependence
of Eq.~\eqref{tunnel1}, the distribution of the computed conductivities was approximately of the log-normal form.

\begin{figure}[t]
\begin{center}
\includegraphics[scale=0.31,clip=true]{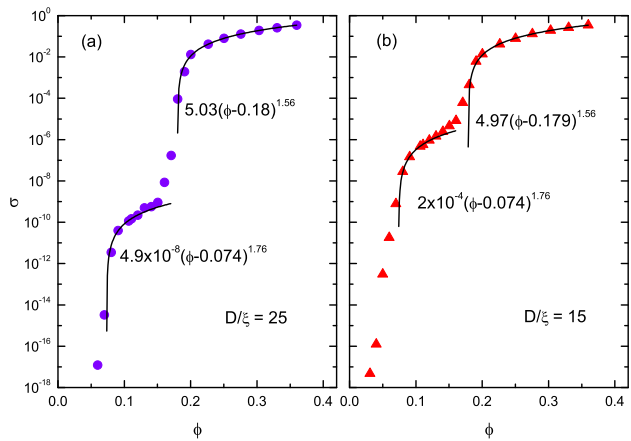}
\caption{(Color online) The lattice Monte Carlo conductivity as a function of $\phi$
for (a) $D/\xi=25$ and (b) $D/\xi=15$.
The solid lines are the best fits with the percolation formula $\sigma=\sigma_0(\phi-\phi_c)^t$. }\label{fig4}
\end{center}
\end{figure}

It is interesting to notice that there is an overall quasi-quantitative
agreement between the EMA and the MC results, for both lattice and continuum models,
meaning that the EMA formulation of Sec.~\ref{EMA} captures well the physics of the problem.
For example, the first percolation threshold at $p_1=1/3$
($\phi_1=\pi p_1/6\simeq 0.175$) obtained from the EMA on the cubic lattice is very close to the
critical value $p_1\simeq 0.3116$ ($\phi_1\simeq 0.163$) for the site percolation problem on the
simple cubic lattice.\cite{Stauffer1994,Sahimi2003,Deng2005}
An important expected difference is however found in the region $\phi>\phi_1$
where the MC conductivity for the lattice case should follow a $(\phi-\phi_1)^t$
dependence with a critical
exponent $t\simeq 2$ instead of the EMA exponent of $t=1$.\cite{Sahimi2003} This is indeed verified in Fig.~\ref{fig4}
where the MC conductivities for $D/\xi=25$ (a) and $D/\xi=15$ (b) are both fitted with $t\simeq 1.56\pm 0.04$,
which is slightly lower than $t\simeq 2$ due to finite size effects
(for the same reason we find $\phi_1\simeq 0.18$ instead of $\phi_1\simeq 0.163$).
Similar critical behaviors are expected in the vicinity of all consecutive percolation thresholds
$\phi_k$ with $k>1$, but due to the limited number of particle densities considered in our MC calculations
and the close proximity of successive $p_k$ values we have been able to fit only the $D/\xi=25$ case
in the vicinity of the second percolation threshold. We have found $t=1.76\pm 0.15$, which agrees within errors
with the exponent at $\phi_1$, and $\phi_2\simeq 0.074$ ($p_2\simeq 0.141$) which is close to
the expected value $\phi_2\simeq 0.072$ ($p_2\simeq 0.137$).\cite{Pike1974}
These same values of $t$ and $\phi_2$ appear to reproduce quite well the percolation behavior around
the second percolation threshold also for the $D/\xi=15$ case [Fig.~\ref{fig4}(b)].

Another important agreement between the EMA and the MC results is the tendency of the lattice model conductivity to
become tunneling-like as $D/\xi$ decreases. This is illustrated in Fig.~\ref{fig3}(c) where the MC conductivities
of Fig.~\ref{fig3}(a) and Fig.~\ref{fig3}(b) are plotted as functions of $\phi^{-1/3}$. The asymptotic regime
for small $\sigma$ of the continuum model (open symbols) follow Eq.~\eqref{EMA9} with the coefficient $0.63$
replaced by $0.7$, as obtained from the critical path approximation applied to a dilute system of
impenetrable spheres.\cite{Seager1974} In complete analogy with the EMA results,
the conductivities of the lattice model
[filled symbols in Fig.~\ref{fig3}(c)] steadily tend to this tunneling regime as $D/\xi$ decreases, and
for $D/\xi=5$ the MC results of the two models are practically indistinguishable for all values of $\phi$.

\section{Percolation-to-hopping crossover}
\label{crossover}

Both EMA and MC results point toward a substantial equivalence between the cubic lattice and the
continuum models for sufficiently small values of $D/\xi$. Given that the two models considered
represent two extreme limits of how the conducting particles may be ideally arranged in
an insulator-conductor composite, such equivalence is important for the understanding of the transport
properties in real composites, whose microstructure is neither completely lattice-like nor
exactly an equilibrium distribution in the continuum. We find it therefore useful to define a
measure for the deviation between the lattice and continuum models in order to follow quantitatively
how these two extremes approach each other as $D/\xi$ decreases.

\begin{figure}[t]
\begin{center}
\includegraphics[scale=0.3,clip=true]{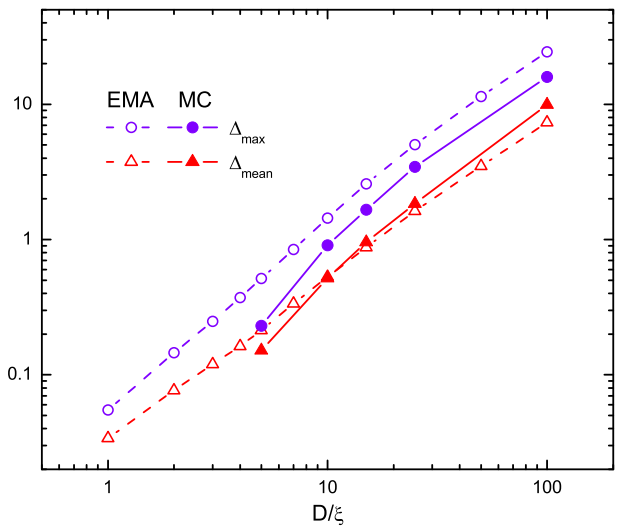}
\caption{(Color online) Maximum value $\Delta_{\rm max}$ and mean value $\Delta_{\rm mean}$
extracted from the whole $\phi$ dependence of Eq.~\eqref{delta} and for several values of $D/\xi$.
The open symbols are the EMA results, while the filled symbols refer to the MC calculations. }\label{fig5}
\end{center}
\end{figure}

To this end, we introduce the following quantity:
\begin{equation}
\label{delta}
\Delta=\left\vert\log\!\left(\frac{\sigma_{\rm latt.}}{\sigma_{\rm cont.}}\right)\right\vert,
\end{equation}
where $\sigma_{\rm latt.}$ and $\sigma_{\rm cont.}$ are the conductivities for the lattice
and the continuum models, respectively, and the $\log$ is the logarithm with base $10$.
Hence, according to Eq.~\eqref{delta}, $\Delta=M$ if for a certain $\phi$ value $\sigma_{\rm latt.}$
differs from $\sigma_{\rm cont.}$ by $M$ orders of magnitude. In Fig.~\ref{fig5}
we plot the maximum value $\Delta_{\rm max}$ and the mean value $\Delta_{\rm mean}$ of $\Delta$
as calculated over the entire range of $\phi$ considered and for several values of $D/\xi$.
The open symbols are the $\Delta_{\rm max}$ and $\Delta_{\rm mean}$ values as extracted from the EMA calculations
while the filled symbols refer to the MC results. As clearly seen in the figure, when $D/\xi$ decreases
both EMA and MC data display a steady decrease of $\Delta_{\rm max}$ and $\Delta_{\rm mean}$.
For $D/\xi < 10$, $\Delta_{\rm max}$ becomes less than unity, which means that $\sigma_{\rm latt.}$
and $\sigma_{\rm cont.}$ differ at most by less than one order of magnitude in the whole range of
$\phi$ values. We also note that for $D/\xi < 10$ the MC results display a somewhat stronger decrease
than the EMA results. This feature does not seems to be due to finite size effects in the MC
calculations, and we attribute it to a real deviation from the EMA results. We can then identify
the value of  $D/\xi\simeq 5$ as a crossover between percolation and hopping regimes below which the conductivities of
the lattice and continuum models differ significantly less than one order of magnitude.

As repeatedly stressed above, the lattice and continuum models are rather idealized
representations of the true microstructure of real conductor-insulator composites. However they also
define two extreme boundaries which delimit somehow all the possible configurations that can be found
in isotropic and homogeneous composites. For this reason, the results of Fig.~\ref{fig5} can be considered
as upper boundaries of more realistic composites, and our $D/\xi\simeq 5$ result probably underestimates
the crossover point between percolation and hopping for composites whose microstructure deviates
from an ideal lattice model.

\section{Discussion and conclusions}
\label{concl}

The results presented in the previous sections consistently show that the GTN model of conductor-insulator
composites, where each particle is connected via tunneling to all others, is capable of explaining the
appearance of both percolation and hopping regimes depending on few characteristics of the composite.
In particular, we have identified the ratio $D/\xi$ and the composite microstructure as the relevant
variables that control the switch from one regime to the other. For large values of $D/\xi$, percolation
arises in composites whose microstructure can approach a regular lattice that is fractionally occupied by
conducting particle, while a hopping regime of the conductivity is always obtained for equilibrium
distributions of conducting particles in the continuum. As $D/\xi$ decreases, the conductivity
of the lattice-like composites gradually loses its percolating character and approaches the hopping regime
which characterizes the continuum limit. For $D/\xi\lesssim 5$ the composite conductivity
displays hopping behavior independent of the specific microstructure and is practically indistinguishable
from that arising from an equilibrium distribution of particles in the continuum.

Given that the values of the tunneling characteristic length $\xi$ do not exceed a few nanometers,
our analysis predicts that composites whose conducting fillers have nanometric sizes should always
display a hopping-like ({\it i. e.}, percolation-less) behavior independent of the particular
distribution of the conducting particles in the composite. Instead, composites with conducting
filler sizes larger than a fraction of microns should display percolation or hopping behaviors
depending on the whether the microstructure is more lattice-like or more continuum-like, respectively.
In view of the above, it appears then reasonable that dilute filler polymer-based
composites will display hopping-like behavior also for filler diameters substantially larger than
$\xi$ ({\it i. e.}, in the order of some hundreds of nanometers) as we have verified recently
in Ref.~\onlinecite{Ambrosetti2010}. At the same time, our theory also explains the conductivity
behavior of composites made of mixtures of hard conducting and insulating particles.
In particular, we note that in dense ensembles of conducting spheres the arrangement in continuum systems
is similar to that of lattices.\cite{Scher1970,Lavrik2001,Kumar2005}
In this respect, the results of Ref.~\onlinecite{Toker2003} on the conductivity of co-sputtered Ni-SiO$_2$
cermets represent a nice example of tunneling-driven percolating behavior on a lattice-like microstructure.
Indeed, the data of Ref.~\onlinecite{Toker2003} display multiple percolation thresholds as the
concentration of conducting Ni grains is reduced, very much like the behavior shown in Fig.~\ref{fig4}
for the MC conductivity of a fractionally occupied lattice model. The difference
of about $3$ orders of magnitude between the conductivity at the largest Ni concentration and that at the
first percolation threshold can be reproduced by our model by setting $D/\xi\simeq 10$ which, by using
the measured mean Ni grain size $D\approx 10$ nm,\cite{Toker2003} leads to $\xi\approx 1$ nm, which is
of the expected order of magnitude (see e.g. Refs.~\onlinecite{Shklovskii1984,Seager1974}) and compares
well with the $\xi$ values extracted from other composites with spherical fillers.\cite{Ambrosetti2009,Ambrosetti2010}
It should be noted however that the lattice model applies only partially to the Ni-SiO$_2$ data
because the nonuniversal value of the conductivity exponent $t$ observed
in the vicinity of the second percolation threshold in Ref.~\onlinecite{Toker2003} cannot be reproduced by
our lattice MC results (see Fig.~\ref{fig4}). As to be discussed elsewhere, a nonuniversal value of
$t$ (in the sense that we have specified in Refs.~\onlinecite{Johner2008,Grimaldi2006}) could be obtained within
the GTN model by allowing a finite dispersion in the distances of the second nearest neighbors.

Before concluding it is worth to point out some limitations of the theory presented above. First, for
simplicity, we have considered composites whose conducting fillers are given by monodispersed spheres.
Even maintaining that for some classes of composites the shape of the fillers can be approximated by a sphere,
the collection of such spheres in a real composite has however some degree of polydispersivity, which can be
large enough to make the fractionally occupied periodic lattice model an inappropriate description of the possible
configuration of the microstructure. However, as stressed above, the lattice model should
be regarded as an extreme deviation from a continuum dispersion of particles (monodispersed or not), so that
the crossover studied in Sec.~\ref{crossover} represents an upper limit also for the case of polydispersed
spherical particles. The second limitation is associated with the fact the we have limited ourselves to the case
of isotropic particle fillers, and we have not attempted to analyze how percolation and hopping behaviors
arise for anisotropic filler shapes such as rodlike or platelike particles. An analysis of these cases
would be interesting also in relation to the appearance of nematic phases for large volume fractions of
anisotropic fillers and to lattice-like arrangements driven by attractive forces in dispersions
of rodlike particles.

\acknowledgements
This work was supported in part by the Israel
Science Foundation (ISF), and in part by the Swiss National Science Foundation (Grant No. 200021-121740).


\begin{thebibliography}{38}
\bibitem{noteW}
Note that, since the function $W(r)$ in Eq.~\eqref{W} has the form of a Fermi distribution with
$\xi/2$ palying the role of temperature, one
has $-dW/dr\rightarrow \delta(r-r^*)$ for $\xi\rightarrow 0$.
\bibitem{noteCPA}
We remind that the CPA amounts to keep only the subset of tunnel conductances $g(r_{ij})$
having $r_{ij}\leq r_c$, where $r_c$ is the largest among the inter-particle distances such
that the so-defined subnetwork forms a conducting percolating cluster. It then follows that
$r_c$ is such that $Z(r_c)=Z_c$ is satisfied, where $Z_c$ is the critical coordination number.
Finally the CPA is obtained by assigning the value $g_c=g_0\exp[-2(r_c-D)/\xi]$ to all the
tunnel conductances of the entire network. In the present EMA version of the CPA, $g_c$ is replaced
by $g^*$, Eq.~\eqref{EMA8}, and the criterion $Z(r_c)=Z_c$ is replaced by Eq.~\eqref{EMA7}.
\end{thebibliography}
\end{document}